\documentclass{emulateapj}
\pdfoutput=1

\nonstopmode 

\usepackage[usenames,dvipsnames,svgnames,table]{xcolor}
\usepackage{enumerate}
\usepackage[T1]{fontenc}
\usepackage{ae,aecompl,amsmath}

\usepackage{amsmath,amstext}
\usepackage[T1]{fontenc}
\usepackage{apjfonts} 

\shorttitle{CGM Fundamental Plane}
\shortauthors{Bordoloi et al.}
\usepackage[colorinlistoftodos]{todonotes}

\providecommand\scription[2]{\scriptsize#1$\;${\scriptsize\uppercase\expandafter{\romannumeral
#2}}\relax}%
\newcommand{\mwlya}{W_{\rm Ly\alpha}}
\newcommand{\wlya}{$\mwlya$}
\providecommand{\Lya}{\ensuremath{\mbox{Ly}\alpha}}
\providecommand{\lya}{\ensuremath{\mbox{Ly}\alpha}}

   \providecommand{\HI}{\ensuremath{\mbox{\ion{H}{1}}}}
\providecommand{\OVI}{\ensuremath{\mbox{\ion{O}{6}}}}
\providecommand{\CIV}{\ensuremath{\mbox{\ion{C}{4}}}}

\def\spose#1{\hbox to 0pt{#1\hss}} \def\simlt{\mathrel{\spose{\lower
3pt\hbox{$\mathchar"218$}}
     \raise 2.0pt\hbox{$\mathchar"13C$}}}
\def\simgt{\mathrel{\spose{\lower 3pt\hbox{$\mathchar"218$}}
     \raise 2.0pt\hbox{$\mathchar"13E$}}}

\begin{document}

\title{On the CGM Fundamental Plane: The Halo Mass Dependency  of Circumgalactic \ion{H}{1}}

\author{Rongmon Bordoloi\altaffilmark{1,2}, 
J. Xavier Prochaska\altaffilmark{3}, 
Jason Tumlinson\altaffilmark{4,5}, 
Jessica K. Werk\altaffilmark{6},
Todd M. Tripp\altaffilmark{7}, 
 \&
Joseph N. Burchett\altaffilmark{3}
}
\altaffiltext{1}{MIT-Kavli Center for Astrophysics and Space Research, 77 Massachusetts Avenue, Cambridge, MA, 02139, USA;\href{mailto:bordoloi@mit.edu}{bordoloi@mit.edu}}
\altaffiltext{2}{Hubble Fellow}
\altaffiltext{3}{UCO/Lick Observatory, University of California, Santa Cruz, CA}
\altaffiltext{4}{Space Telescope Science Institute, Baltimore, MD}
\altaffiltext{5}{Department of Physics \& Astronomy, Johns Hopkins University, Baltimore, MD}
\altaffiltext{6}{University of Washington, Department of Astronomy, Seattle, WA, USA}
\altaffiltext{7}{Department of Astronomy, University of Massachusetts, Amherst, MA}

\begin{abstract}
We analyze the equivalent widths of  \ion{H}{1} {\lya} ({\wlya})  absorption from the inner ($R < 160$ kpc) circumgalactic medium (CGM) of 85 galaxies at $z \sim 0$ with stellar masses $M*$ ranging $\rm{8 \leq log M* / M_{\odot} \leq 11.6}$. Across three orders of magnitude in stellar mass, the CGM of present-day galaxies exhibits a very high covering fraction of cool hydrogen gas ($f_C = 87\pm 4$\%) indicating that the CGM is ubiquitous in modern, isolated galaxies. When  \ion{H}{1} {\lya} is detected, its equivalent width declines with increasing radius regardless of the galaxy mass, but the scatter in this trend correlates  closely with $M*$. Using the radial and stellar mass correlations, we construct a planar surface describing the cool CGM of modern galaxies: $\log W^{\rm{s}}_{HI 1215} \; = \;  (0.34 \pm 0.02) -( 0.0026 \pm 0.0005)\times (R) + (0.286 \pm 0.002) \times  \log (M*/M_{\odot})$. The RMS scatter around this bivariate relation is $\sim$0.2 dex. We interpret the explicit correlation between \wlya\ and $M*$ to arise from the underlying dark matter halo mass ($M_{halo}$), thereby suggesting a CGM fundamental plane between {\wlya}, $R$ and $M_{halo}$. This correlation can be used to estimate the underlying dark matter halo mass  from observations of saturated \ion{H}{1} {\lya}  in the CGM of a modern galaxy.

\end{abstract}

\keywords{galaxies: halos --- quasars: absorption lines --- intergalactic medium}

\section{Introduction}

In the original $\Lambda$CDM paradigm, gas accreting onto dark matter halos cools, sinks to the center of the halo, and condenses to form stars \citep{White1978,Mo1996}. Advances in multi-wavelength observations over the last two decades have resulted in a more sophisticated picture of galaxy evolution in which gas and metals cycle into and out of galaxies continually over time. Gas accretion fuels star-formation and grows the galaxies. Feedback from supernovae or active galactic nuclei (AGN) eject gas and metals from galaxies and help to quench star-formation (e.g., \citealt{Somerville2015}). This baryon cycle takes place in the region surrounding galaxies called the circumgalactic medium (CGM) \citep{Chen2010a,Steidel2010,Tumlinson2011a,Bordoloi2011,Bordoloi2014c,Werk2013,Nielsen2013,Prochaska2017,Tumlinson2017}.

The dominant gravitational potential of dark matter influences halo gas dynamics: it may set the coronal gas temperature, the pressure gradient, and the cooling and cloud-infall timescales \citep{Mo1996}. Observations of CGM gas  provide an opportunity to study gas dynamics within halos and to understand these influences. 

The dependence of metal-line strength on halo and/or stellar mass has been demonstrated with Mg~II ions \citep{Chen2010b,Bordoloi2011,Churchill2013}. However, Mg~II absorption is also believed to trace gas inflows and outflows up to $\sim$100 kpc \citep{Bordoloi2011,Kacprzak2012,Bouche2013,Bordoloi2014d}.  
Strong Mg~II is rarely seen at large impact parameters and so cannot be used to contrain halo dynamics further out. Most other saturated metal absorption lines are generally undetected at available limits at $R > R_{vir}/2$ \citep{Bordoloi2014c,Liang2014,Johnson2015}. In this work, we investigate the dependence of {\Lya} absorption on host galaxy mass. Because strong {\Lya} absorption is observed even beyond the virial radius, it reveals gas properties across the entire halo.

Pioneering work with the \textit{Hubble Space Telescope} (HST) mapped the projected radial profile ($R$) of {\Lya} and other metal lines around low-$z$ galaxies 
\citep[e.g.,][]{Bowen1995,Lanzetta1995,Morris1993}. These studies have found that the multiphase CGM is a complex and rich environment  with multiple density and temperature phases \citep{Tripp2011, Chen2001, Bowen2002, Tripp2008,Prochaska2011}. In recent years, large galaxy surveys combined with sensitive QSO spectra from the \textit{Cosmic Origin Spectrograph} have
enabled systematic mapping of the CGM around select samples of galaxies \citep{Werk2012,Tumlinson2013,Bordoloi2014c,Liang2014,Borthakur2015,Burchett2016,Lehner2015,Werk2016, Bordoloi17,Heckman2017}.

In previous work, with the COS-Halos and COS-Dwarfs surveys, our group has examined the CGM around galaxies at $R < 160$\,kpc in a wide range of diagnostic ions. The COS-Halos survey \citep{Tumlinson2013} studied the multiphase CGM around 44 galaxies at $z \sim 0.2$ with $\log M*/M_{\odot} \geq 10$ \citep{Tumlinson2011a,Werk2012,Werk2013}. The COS-Dwarfs survey \citep{Bordoloi2014c} studied the multiphase CGM of 43 galaxies at $z \leq 0.1$ with $8 \leq \log M*/M_{\odot} \leq 10$. In this paper, we combine these datasets to gain a view of cool gas traced by {\Lya} absorption around galaxies spanning three decades in stellar mass. Throughout this work, we adopted a $\Lambda$CDM cosmology  ($\rm{\Omega_{m}}$ = 0.238, $\rm{\Omega_{\Lambda}}$ = 0.762, $\rm{H_{0}}$ = 73.2 $\rm{kms^{-1}\; Mpc^{-1}}$, $\rm{\Omega_{b}}$ = 0.0416). 

\section{Description of the Sample}

Detailed descriptions of survey design and sample selection can be found for COS-Halos in \cite{Tumlinson2013} and for COS-Dwarfs in \cite{Bordoloi2014c}. In short, both surveys utilized HST/COS to acquire UV spectroscopy of bright background quasars spanning 1140-1750\AA\ at R $\sim$ 20,000 and S/N $\sim$ 10. Both these CGM surveys were designed as  ``galaxy selected''  surveys, where the projected galaxy-quasar pairs are selected based on projected physical separation and known foreground galaxy properties and without any {\it a-priori} knowledge of gas properties along the line of sight. 

The COS-Halos survey was optimized to study highly-ionized CGM gas using {\OVI} and other metal line diagnostics, targeting $\sim L^*$ galaxies.  
The COS-Dwarfs survey was designed to target lower mass galaxies  ($\rm{8 \leq log M*/M_{\odot} \leq10}$) and was optimized to study moderately ionized gas ({\CIV}) and other low ion diagnostics \citep{Bordoloi2014c}. 
Both the COS-Dwarfs and COS-Halos surveys target lines of sight which are within $\sim$ 160 kpc from the host galaxies. Owing to the different mass ranges probed, this impact parameter corresponds to $\sim 0.6 R_{\rm vir}$ for COS-Halos and $1 \, R_{\rm vir}$ for COS-Dwarfs. Both surveys cover the HI {\Lya} line, which traces both neutral and predominantly ionized gas. The median statistical 1$\sigma$ uncertainty for {\Lya} equivalent widths 
({\wlya}) is $\approx$ 25 m{\AA}. Using their specific star formation rates (sSFR), we divide these galaxies into star-forming ($\log sSFR \leq -10.6$)  and passive ($\log sSFR > -10.6$), respectively. 

Both the COS-Halos and COS-Dwarfs surveys were designed to target predominantly isolated galaxies \citep{Werk2013,Tumlinson2013,Bordoloi2014c}. In the COS-Halos survey, the galaxies were selected to not have any accompanying galaxies within 1 Mpc of the candidate galaxies \citep{Tumlinson2013}. In COS-Dwarfs, candidate galaxies are selected with SDSS spectroscopic information to have no other galaxies with known redshifts within 300 kpc from the candidate galaxy \citep{Bordoloi2014c}, although a sub-sample of the galaxies have other galaxies at similar redshifts but at larger separations ($0.3-1$\,Mpc). Hence, this study primarily focuses on the inner CGM properties of isolated galaxies and any effects of galactic environment on the CGM gas are not the focus of this study (see, e.g., \citealt{Burchett2016} for the impacts of galactic environment).

\begin{figure*}[!t] 
\begin{center} 
\includegraphics[height=7.5cm,width=8.cm]{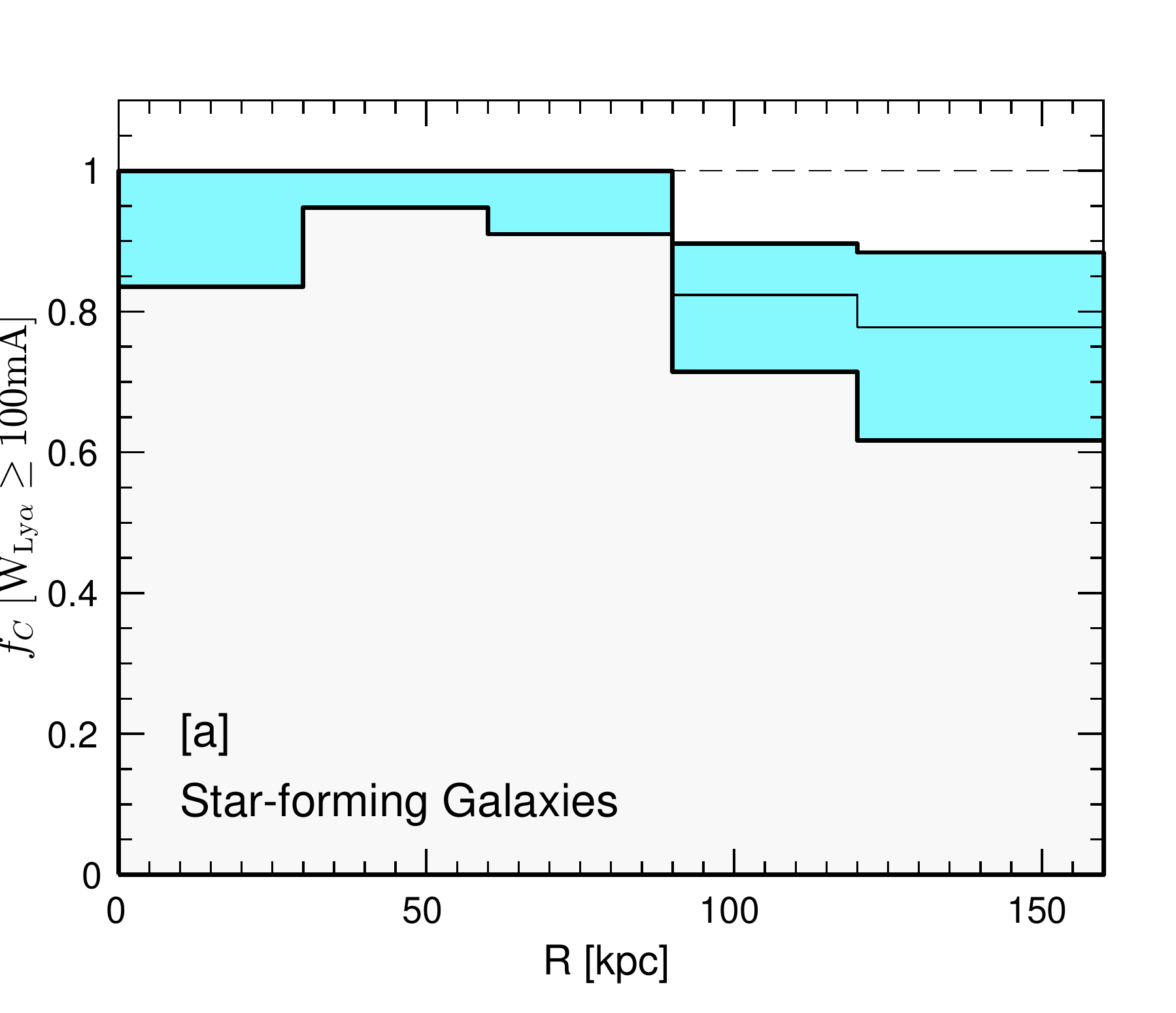}
\includegraphics[height=7.5cm,width=8.cm]{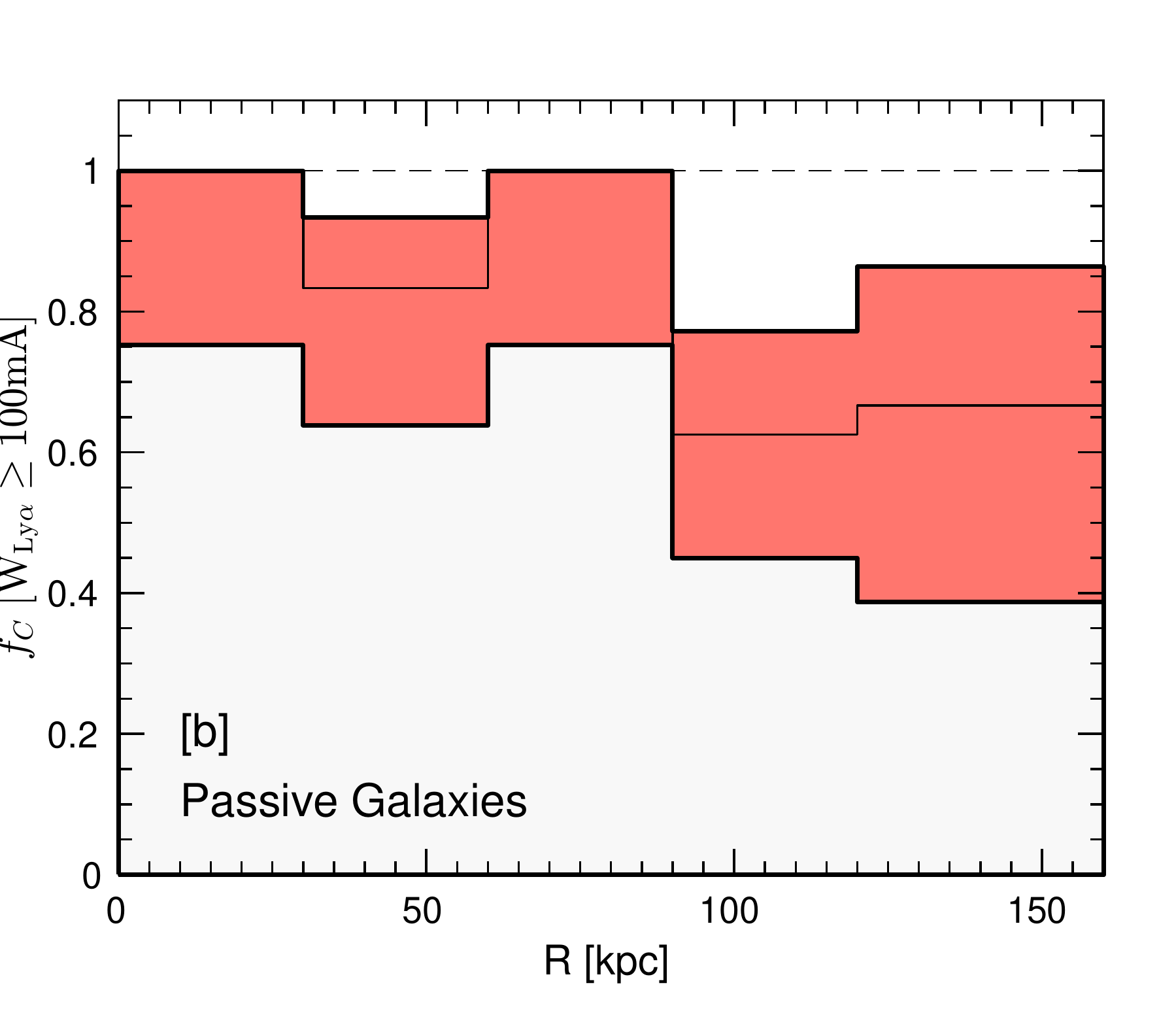}
\includegraphics[height=7.5cm,width=8.cm]{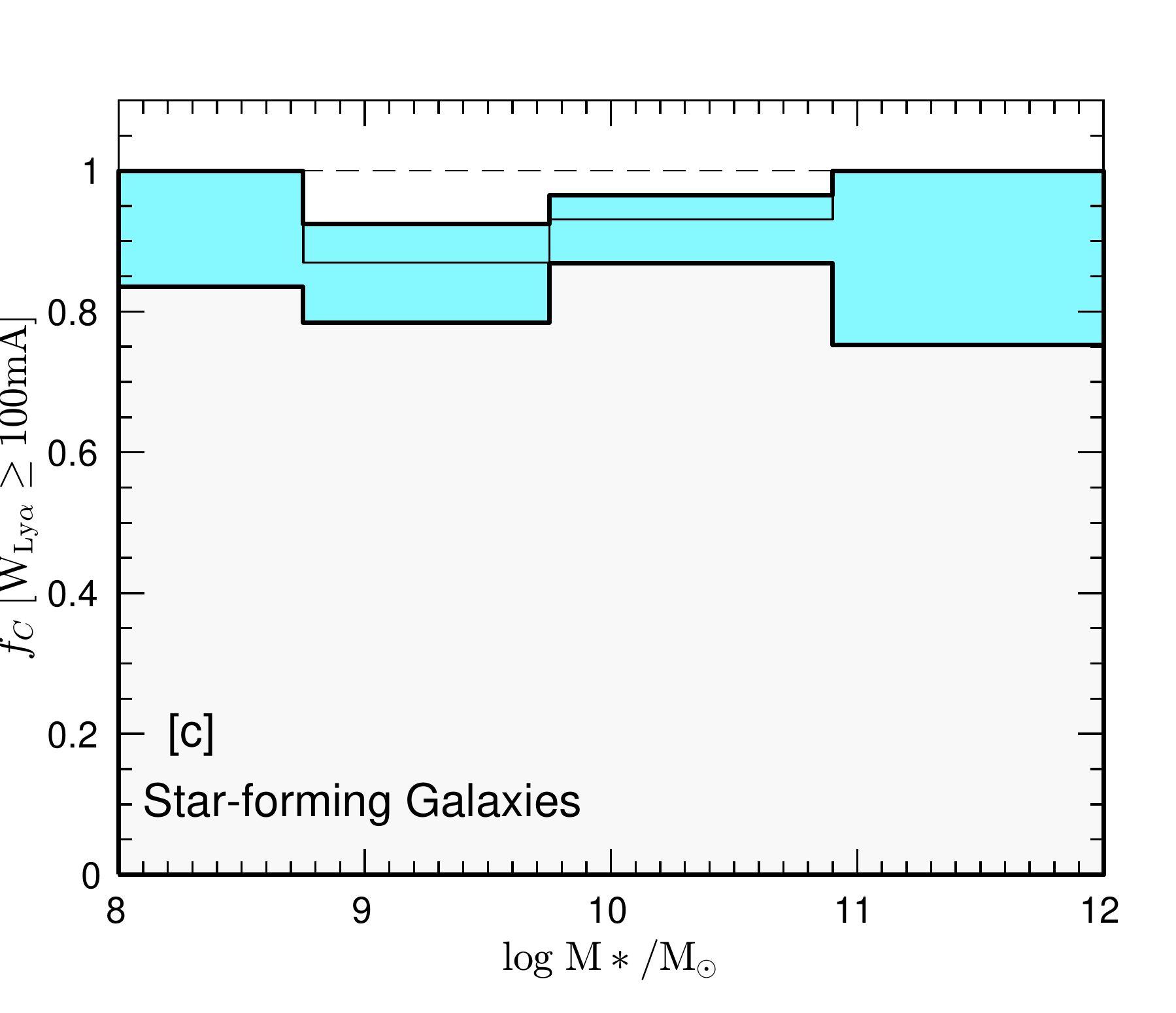}
\includegraphics[height=7.5cm,width=8.cm]{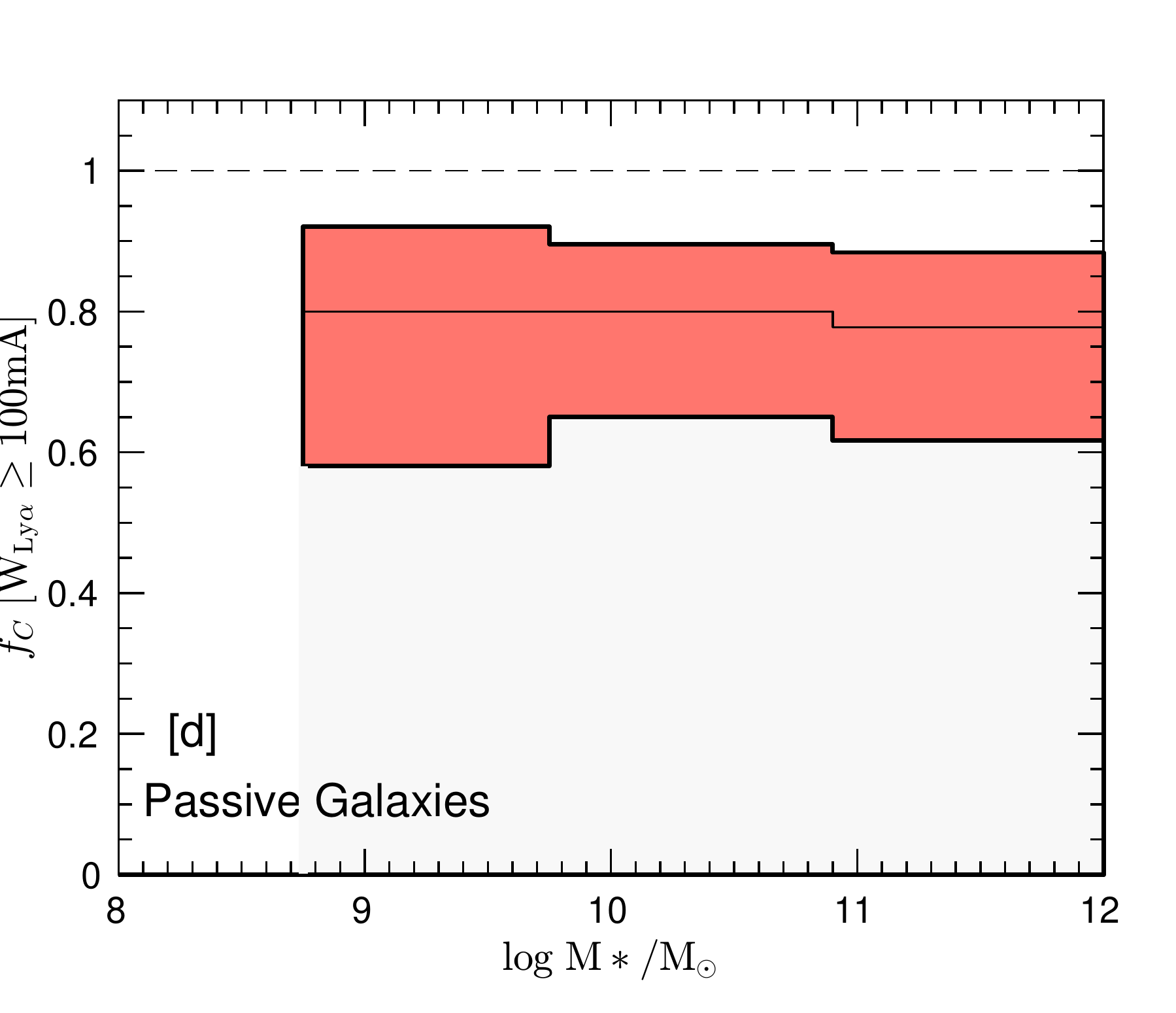}
\label{fig:covering_fraction}
\end{center} 
\caption{{\Lya} covering fraction ($\mwlya \geq 100$ m{\AA}) as a function of observed impact parameter (top panels) and galaxy stellar mass (bottom panels), respectively. Both the star-forming (left panels, blue shaded region) and passive (right panels, red shaded region) galaxies show high incidence of {\Lya} absorption. The width of the shaded regions represents Wilson Score 68\% confidence interval on $f_C$. The bins along x-axes were chosen to have approximately similar number of galaxies per bin.} 
\end{figure*}

\section{Results}

To map the spatial extent of {\Lya} absorption around galaxies, we first investigate the {\Lya} covering fraction ($f_C$) at different impact parameters around galaxies. We define covering fraction as

\begin{equation*}
f_C  =   \frac{N_{W \geq W_{cut}} }{ N_{tot} } \;\; ,
\end{equation*}
\noindent
where $N_{W \geq W_{cut}} $  gives the  number of galaxies within either an impact parameter or a stellar mass bin, which have associated {\Lya} equivalent width ({\wlya}) greater than some cutoff equivalent width $W_{cut}$. $N_{tot} $ gives the total number of galaxies in that same bin. Figure \ref{fig:covering_fraction} shows $f_C$ as a function of impact parameter (panels [a], [b]) and stellar mass (panels [c], [d]) for 
$W_{cut} \geq 100$ m{\AA}. The binomial confidence intervals on $f_C$ are reported as Wilson Score 68\% confidence intervals.
 
Throughout this paper, we adopt an equivalent width cutoff of $W_{cut} \geq 100$ m{\AA}  for $f_C$ measurements. Figure \ref{fig:covering_fraction} shows that {\Lya} absorption is nearly ubiquitous around both star-forming and passive galaxies out to 160 kpc. Of 85 galaxies, 74 exhibit {\Lya} absorption above this cutoff, yielding $f_C$ of 87$\pm$4\% for the full sample. Star-forming galaxies exhibit $f_C$ of 92$\pm$4\% (55/60) within 160 kpc, and $f_C$ of 100\% (33/33) within 90 kpc.  The passive galaxies also exhibit a large $f_C$ of 76$\pm$ 8\% (19/25) within 160 kpc. For $R<90$\,kpc, the passive galaxies show a detection rate of 92$\pm$8\% (11/12).

Figure \ref{fig:covering_fraction}, panels [c] and [d], show the {\Lya} covering fraction as a function of galaxy stellar mass. For the full sample, $f_C$ does not depend on the stellar mass of its host galaxy: 29 out of 33 galaxies at $\log M*/M_{\odot} < 9.75 $ exhibit {\Lya} absorption with $f_C$ of 88$\pm$6\%, and at $\log M*/M_{\odot} \geq 9.75$, $f_C$ = 87$\pm$5\% (45/52). Among star-forming galaxies at $\log M*/M_{\odot} < 9.75$, 
$f_C = 89\pm$6\% (25/28), and at $\log M*/M_{\odot} \geq 9.75$, $f_C=94\pm$4\% (30/32). 
For passive galaxies at $\log M*/M_{\odot} < 9.75$ and $\log M*/M_{\odot} \geq 9.75$, $f_C = 80\pm$17\% (4/5) and 75$\pm$10\% (15/20), respectively. We conclude that in the CGM of galaxies spanning 3 decades in stellar mass ($\rm{8 \leq log M*/M_{\odot} \leq11.6}$), {\Lya} absorption is seen with high incidence for all galaxies. In passive galaxies, its presence at high incidence is a puzzle in light of these galaxies' low or non-existent star formation \citep{Thom2012}.

\begin{figure*}[!t] 
\begin{center} 
\includegraphics[height=7.25cm,width=8.75cm]{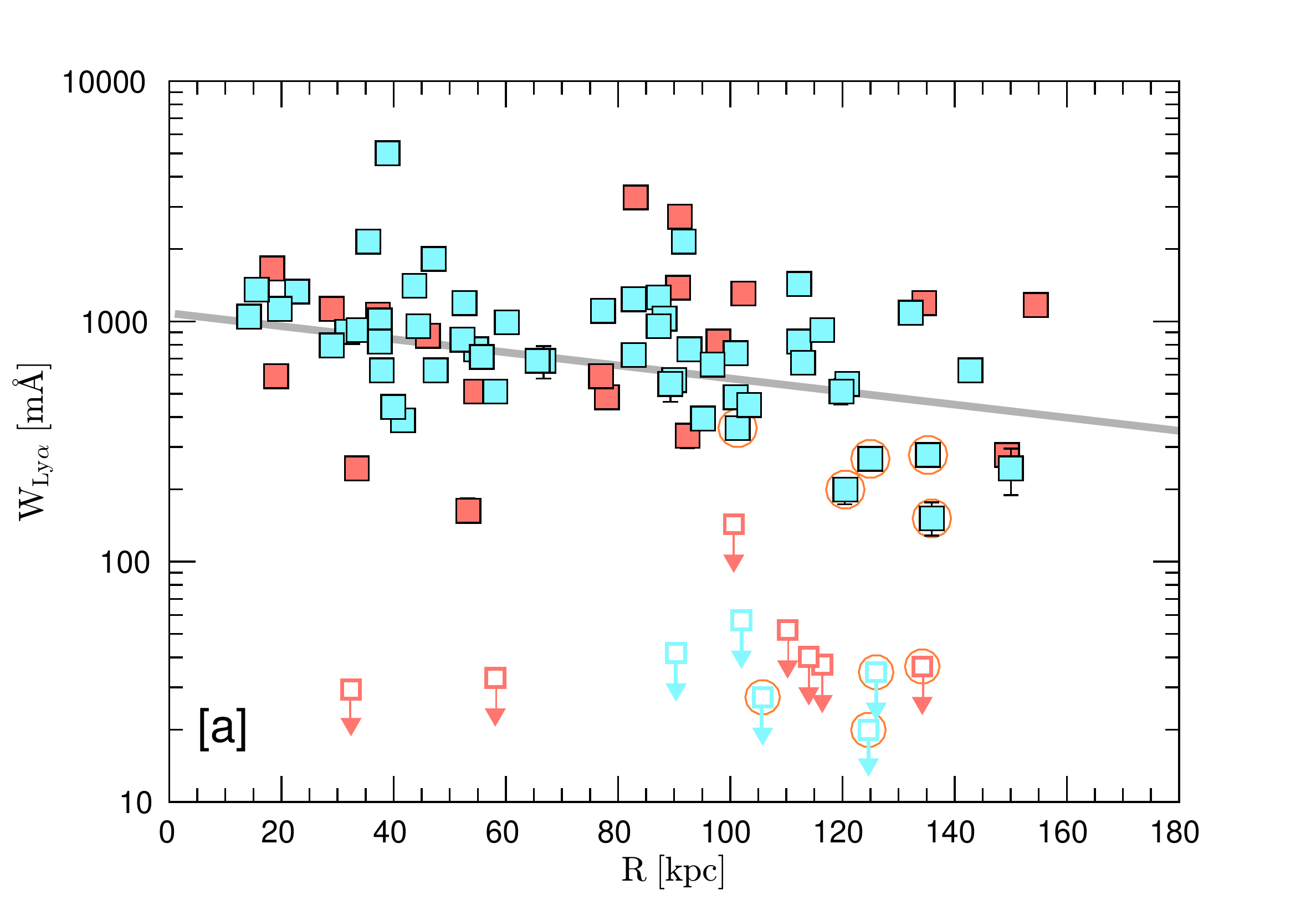}
\includegraphics[height=7.25cm,width=8.75cm]{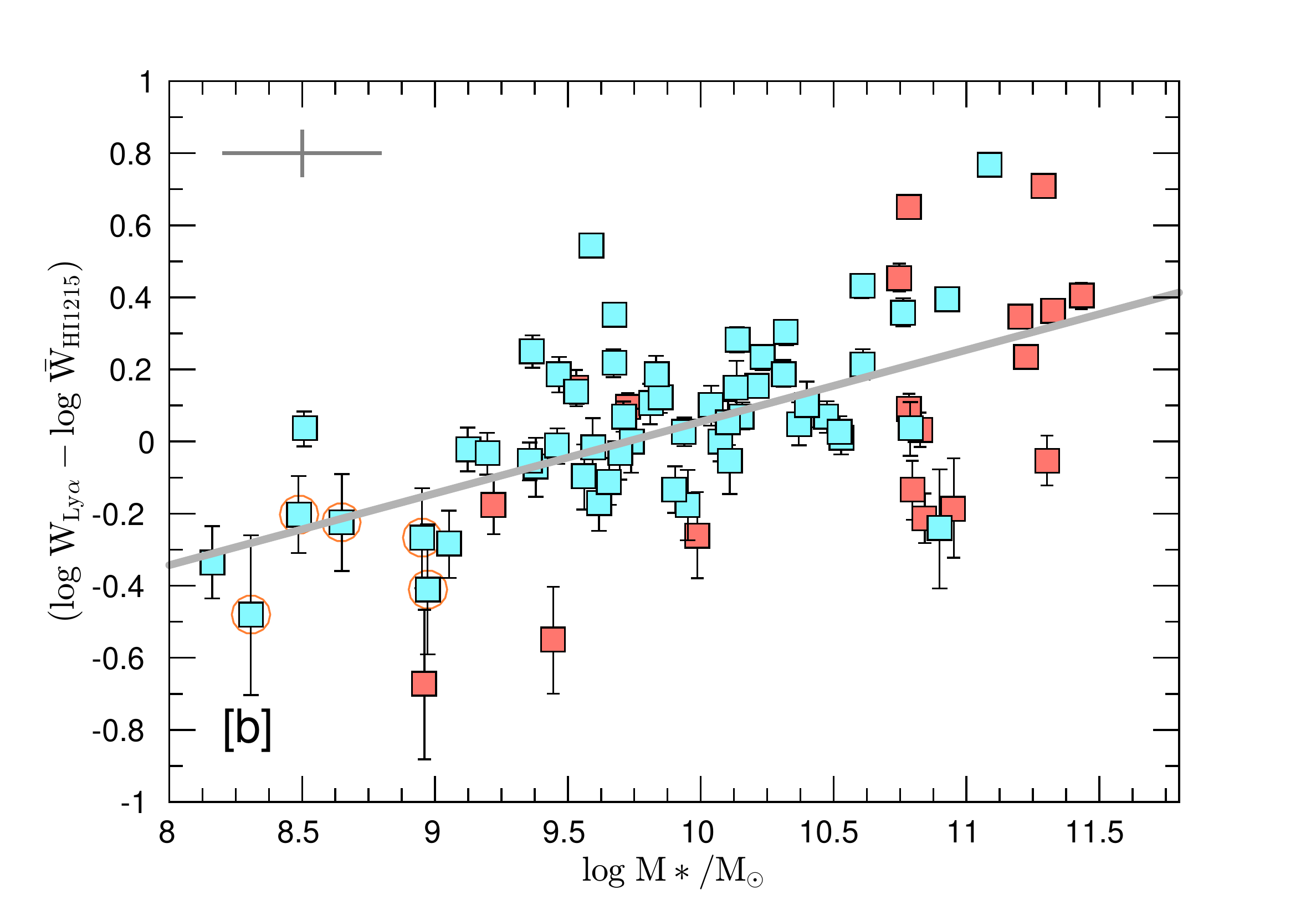}
\includegraphics[height=7.25cm,width=8.75cm]{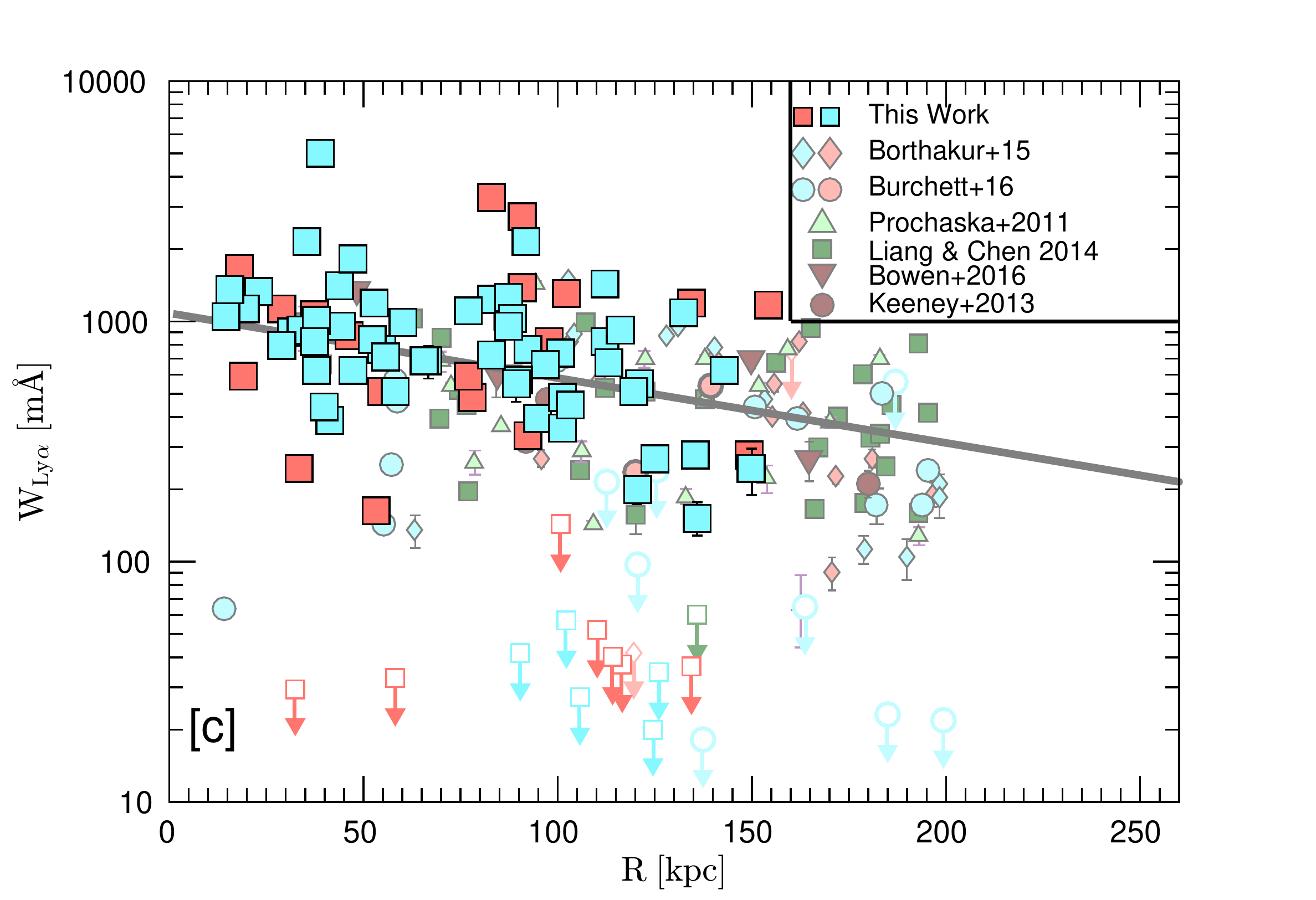}
\includegraphics[height=7.25cm,width=8.75cm]{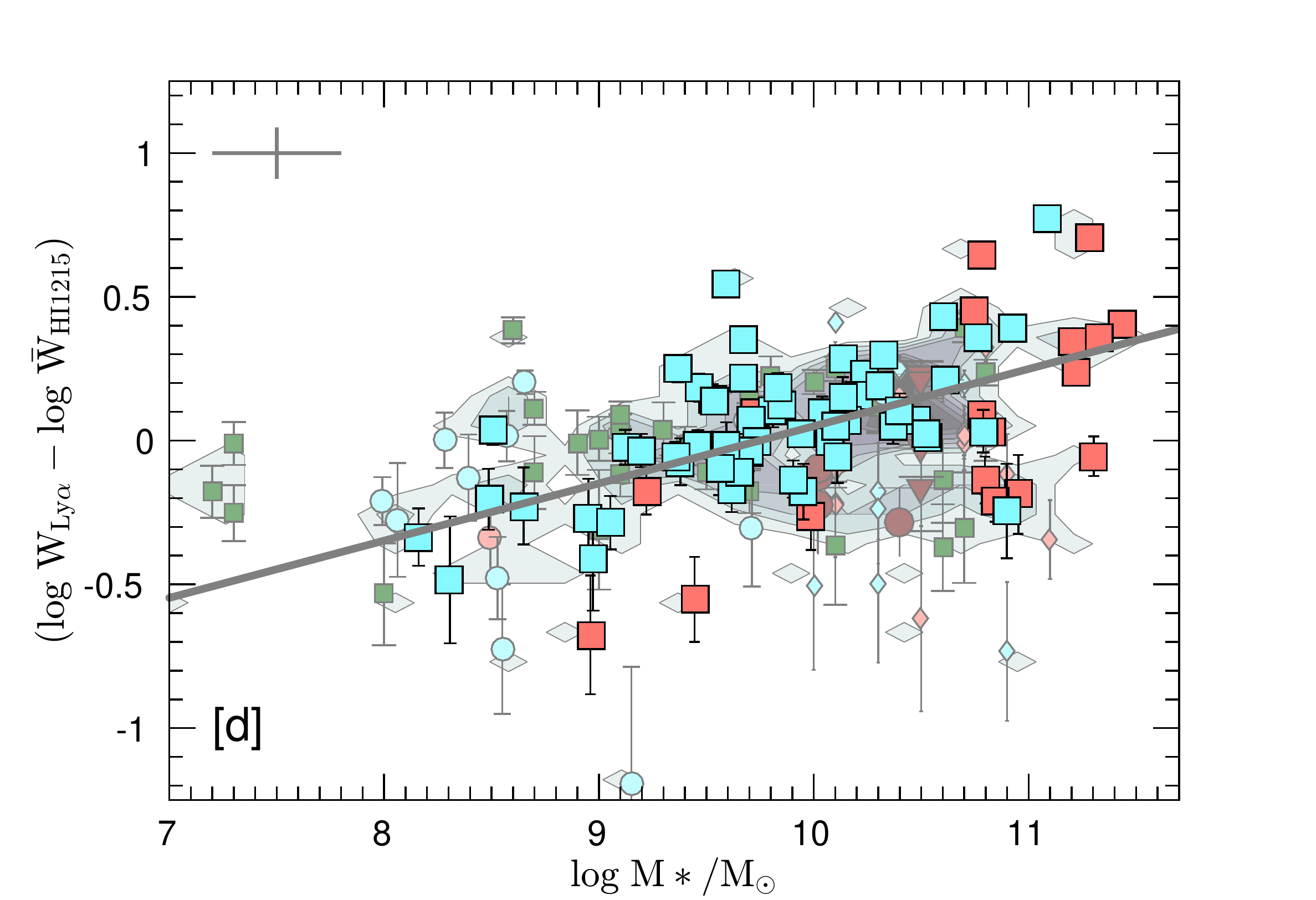}
\label{fig:radial_profile}
\end{center} 
\caption{\textit{Left Panels}: {\Lya} absorption radial profile from this work (panel [a]), and archival data (panel [b]), respectively. The solid gray line shows the best fit radial profile from equation \ref{eqn:radial_profile}. The open yellow circles in panel [a] mark lines of sight at high $R_{vir}$ (see text). \textit{Right Panels}: The {\wlya} absorption residuals as a function of stellar mass, showing strong correlation between the {\Lya} absorption strength and stellar mass for both the star-forming (blue squares) and passive (red squares) galaxies. The underlying contour on the bottom right panel traces the number density of galaxies in the plot. The gray crosses on right panels represent typical uncertainties associated with {\wlya} residual measurements and stellar mass estimates.
} 
\end{figure*}

We further investigate variations in {\Lya} absorption strength as a function of impact parameter. Panel [a] of Figure \ref{fig:radial_profile} shows that the {\Lya} absorption strength decreases slightly at higher impact parameters. The observed {\Lya} equivalent widths around star-forming galaxies (blue filled squares) and passive galaxies (red filled squares) are shown respectively. Most of the measured \wlya\  uncertainties are smaller than the size of the data points. The open squares with arrows show the 2$\sigma$ upper limits 
for non-detections. 
There is a weak anti-correlation between {\Lya} equivalent width and impact parameter. A Kendall's Tau test shows that $\log W_{Ly\alpha} $ and impact parameter are anti-correlated with a P value $\sim$ 0.005; 
this corresponds to ruling out the null hypothesis that there is no correlation between these two quantities at 2.8$\sigma$ significance. Several studies have shown that the {\Lya} equivalent width falls off steeply beyond an impact parameter of $\sim$ 300 kpc \citep{Wakker09,Prochaska2011,Johnson2015}. We parameterize this 1D radial profile as a power law. 
Because the detection rate for {\Lya} absorption is very high, we perform this 
fit only on the detections. 
This parameterization is given by
 \begin{equation}
 \label{eqn:radial_profile}
\log \bar{W}_{HI 1215} \; = \; 3.0343 \pm 0.07 
-  (0.0027 \pm  0.0023) \times  \rm{R} 
 \end{equation}
and is shown as the solid gray line in Figure \ref{fig:radial_profile}, panel [a]. This radial profile is valid even as we add more observations within 200 kpc from other surveys (panel [c], Figure~\ref{fig:radial_profile})
\citep{Borthakur2015,Burchett2016,Prochaska2011,Liang2014,Bowen2016, Keeney2013}. The RMS scatter around this purely radial trend is $\sim$ 0.3 dex.

We now investigate the behavior of the residual {\Lya}  equivalent width once this 1D radial fit is applied to the data. Here the dependent variable is $\log W_{Ly\alpha} - \log \bar{W}_{HI 1215}$, the individual galaxy's {\Lya} equivalent width minus the fitted value from Equation~\ref{eqn:radial_profile}. As there are very few non-detections, from this point onward in the paper, we will focus only on the detections. The top right panel of Figure \ref{fig:radial_profile} shows that these residuals increase steadily with a clear trend with increasing stellar mass, for both star-forming and passive galaxies. That is, more massive galaxies generally lie above the fitted radial trend in Equation \ref{eqn:radial_profile}, while less massive galaxies lie below it. 

A generalized Kendall's tau test can assess the statistical significance of this correlation of residual \wlya\ with stellar mass. We consider all galaxies with detected {\Lya} absorption and obtain significance (P) =  5.02$\times10^{-6}$. This low value excludes the null hypothesis that there is no correlation between \wlya\ residual and host galaxy stellar mass at 4.6$\sigma$ confidence.

We have fit a line to characterize this dependence: 
\begin{equation}
 \label{eqn:mass_profile}
 \begin{split}
\log {W_{Ly\alpha}} - \log \bar{W}_{HI 1215} \; = \;  (-1.94 \pm 0.34) \\
+ (0.199 \pm 0.03) \log (M_{*}/M_{\odot}) .
\end{split}
\end{equation}
This fit is plotted as the gray solid line in the right panels of Figure \ref{fig:radial_profile}. 
The correlation is observed even as we add the literature data
shown in the bottom panels of Figure \ref{fig:radial_profile}. We note that some sightlines around lower mass COS-Dwarfs galaxies probe impact parameters which correspond to higher $R_{\rm vir}$ as compared to COS-Halos sightlines. The COS-Halos sample extends out to an impact
parameter of of $0.62R_{\rm vir}$ while there are 9 sightlines in 
COS-Dwarfs sample with $R > 0.62 R_{\rm vir}$. 
These are marked with open yellow circles in Figure \ref{fig:radial_profile}.  
To test that these sightlines are not biasing the observed correlation, we performed the residual analysis after excluding these 9 objects. This reduces the detected sample size from 74 to 69 galaxies. Not surprisingly, most of these objects have lower mass, because these galaxies would 
have smaller virial radii at the same impact parameter as the high mass systems.

We perform a generalized Kendall's Tau test to assess the statistical significance of the correlation between stellar mass and residual \wlya\ , after removing these high $R_{\rm vir}$ sightlines. We find that a significant correlation exists between stellar mass and residual \wlya\ with a P value = 4.4$\times10^{-4}$. This low value allows us to exclude the null hypothesis that there is no correlation between \wlya\ residual and host galaxy stellar mass at $>$ 3.5$\sigma$ confidence. We also searched for any possible trends of  {\Lya} residual equivalent widths with the sSFR of the host galaxies and found no statistically significant trend (P = 0.93). 
 
\begin{figure*}[!t] 
\begin{center} 
\includegraphics[height=7.5cm,width=8.75cm]{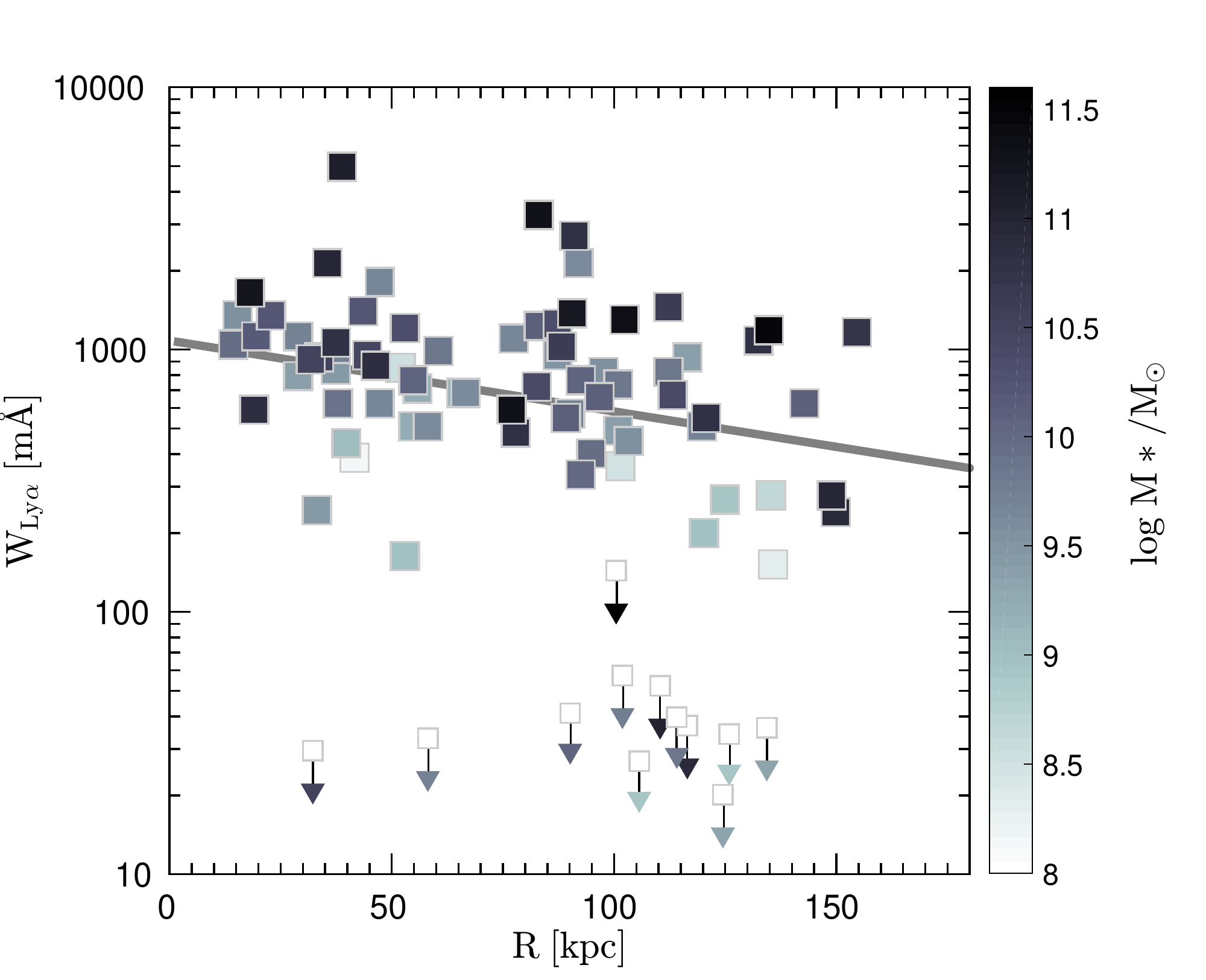}
\includegraphics[height=7.5cm,width=8.75cm]{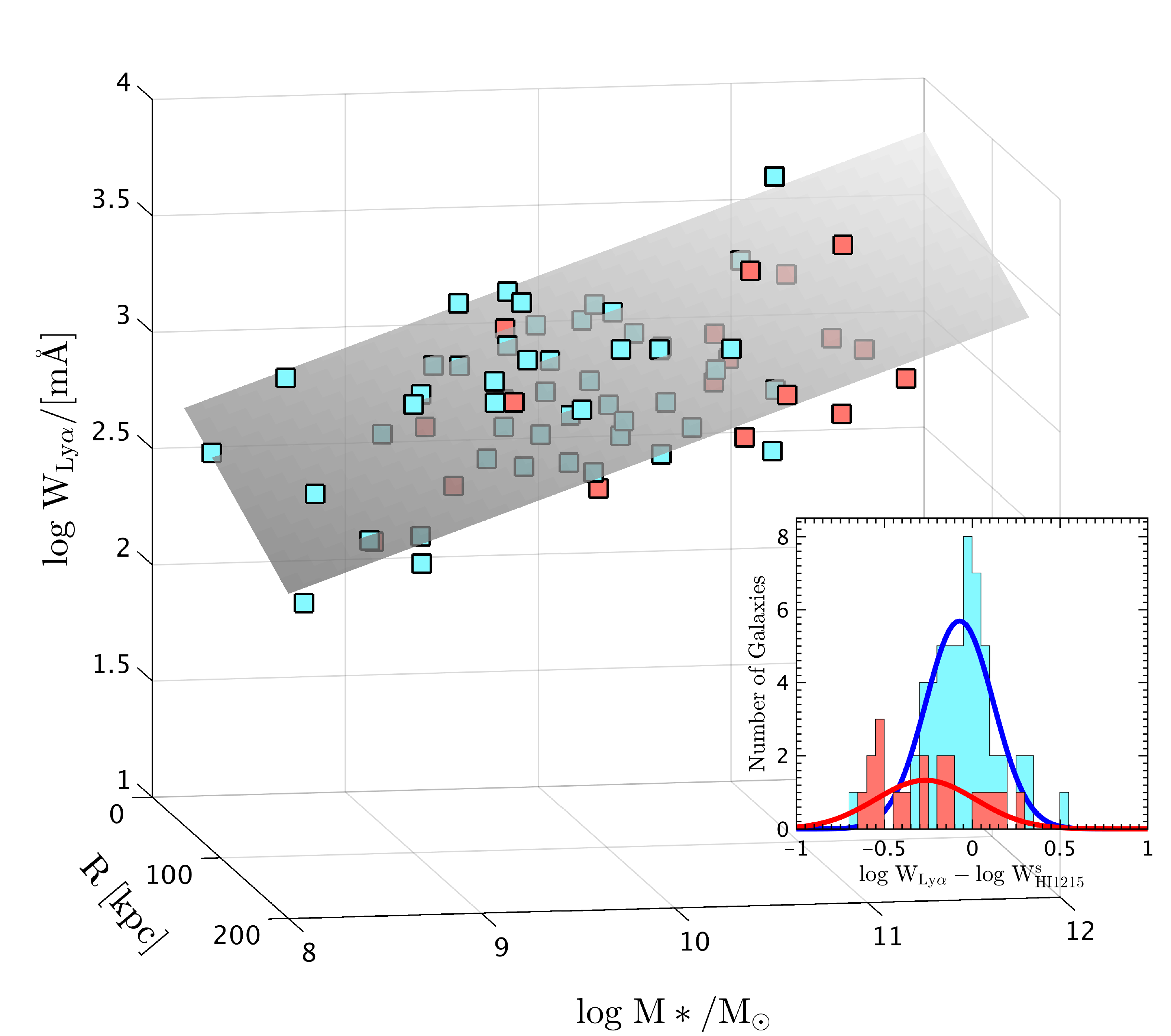}
\label{fig:eqw_surf}
\end{center} 
\caption{ \textit{Left Panel:} The radial profile of {\Lya} absorption for the same galaxies, color coded with their stellar mass. The higher mass galaxies typically segregate towards higher equivalent widths at all impact parameters. Both panels are showing data from this work only. \textit{Right Panel:}A surface fit showing the {\Lya} absorption within 160 kpc of the host galaxies across three decades of stellar mass. At a given stellar mass and impact parameter, the {\Lya} absorption strength can be predicted by the slope of this surface. The inset shows the residual equivalent widths of the fit for the whole sample.} 
\end{figure*}

The mass dependence of \wlya\ can be better visualized in Figure \ref{fig:eqw_surf} (left panel), which shows the {\Lya} radial absorption profile color-coded by their stellar mass. The high-mass galaxies trend towards higher equivalent widths at all impact parameters, visually indicating a dependence with stellar mass.

This analysis suggests that \wlya\ can be described primarily as a function of two variables: the impact parameter and the stellar mass of the host galaxy. We quantify this dependence by fitting a surface to the {\Lya} absorption strength as a function of stellar mass and impact parameter, given as 
\begin{equation}
 \label{eqn:surf_profile}
 \begin{split}
 \log W^{\rm{s}}_{HI 1215} \; = \;  (0.34 \pm 0.02) -( 0.0026 \pm 0.0005)*(R) \\
 +  (0.286 \pm 0.002)* \log (M*/M_{\odot}) .
 \end{split}
 \end{equation}

Figure \ref{fig:eqw_surf} (right panel) shows the surface fit to the COS-Halos and COS-Dwarfs data. The inset shows the residual {\Lya} equivalent widths for star-forming (blue) and passive (red) galaxies respectively. 
The surface fit characterizes well the {\Lya} absorption around 
star-forming galaxies, yielding an RMS scatter $\sim$ $0.18$ dex. The inset of Figure \ref{fig:eqw_surf} (right panel) shows this distribution of residuals around the fitted 2D plane. Equation \ref{eqn:surf_profile} is used for both star-forming and passive galaxies and can describe the {\Lya} absorption strength around galaxies spanning a mass range of  $\rm{8 \leq log M*/M_{\odot} \leq11.6}$. For passive galaxies, this surface fit shows a modestly higher scatter (0.29 dex). The slightly higher scatter could be caused by two different factors. First, higher mass passive galaxies are in general seen in over-dense regions and might be tracing group environments. Hence there might be other unforeseen environmental effects which increase the scatter \citep{Burchett2017}. Second, it is possible that the slope of the surface fit is different for star-forming and passive galaxies. Since our sample is dominated by star-forming galaxies, they drive the fit in equation \ref{eqn:surf_profile}. This could be another reason for the increased scatter in the residual {\Lya} equivalent widths of passive galaxies. The sample size of passive galaxies in this work is not adequate to explore these. We will analyze this issue in a future study.

\section{The dynamics of the cool CGM gas}

Because the observed {\Lya} absorption of the inner CGM is primarily saturated (i.e.,\ the peak optical depth $\tau_{\rm Ly\alpha} \gg 1$), 
the \wlya\ values are likely a better measure of bulk kinematics rather than the column density of the gas. The velocity centroids and ranges of {\Lya} absorbers in the COS-Halos and COS-Dwarfs surveys are generally consistent with being bound to their host dark matter halo even when projection effects are accounted for (\citealt{Tumlinson2013}, Bordoloi et al. in prep), but this does not by itself explain the scaling with $R$ and $M*$ observed here. 

\begin{figure*}[!t] 
\begin{center} 
\includegraphics[scale=0.435]{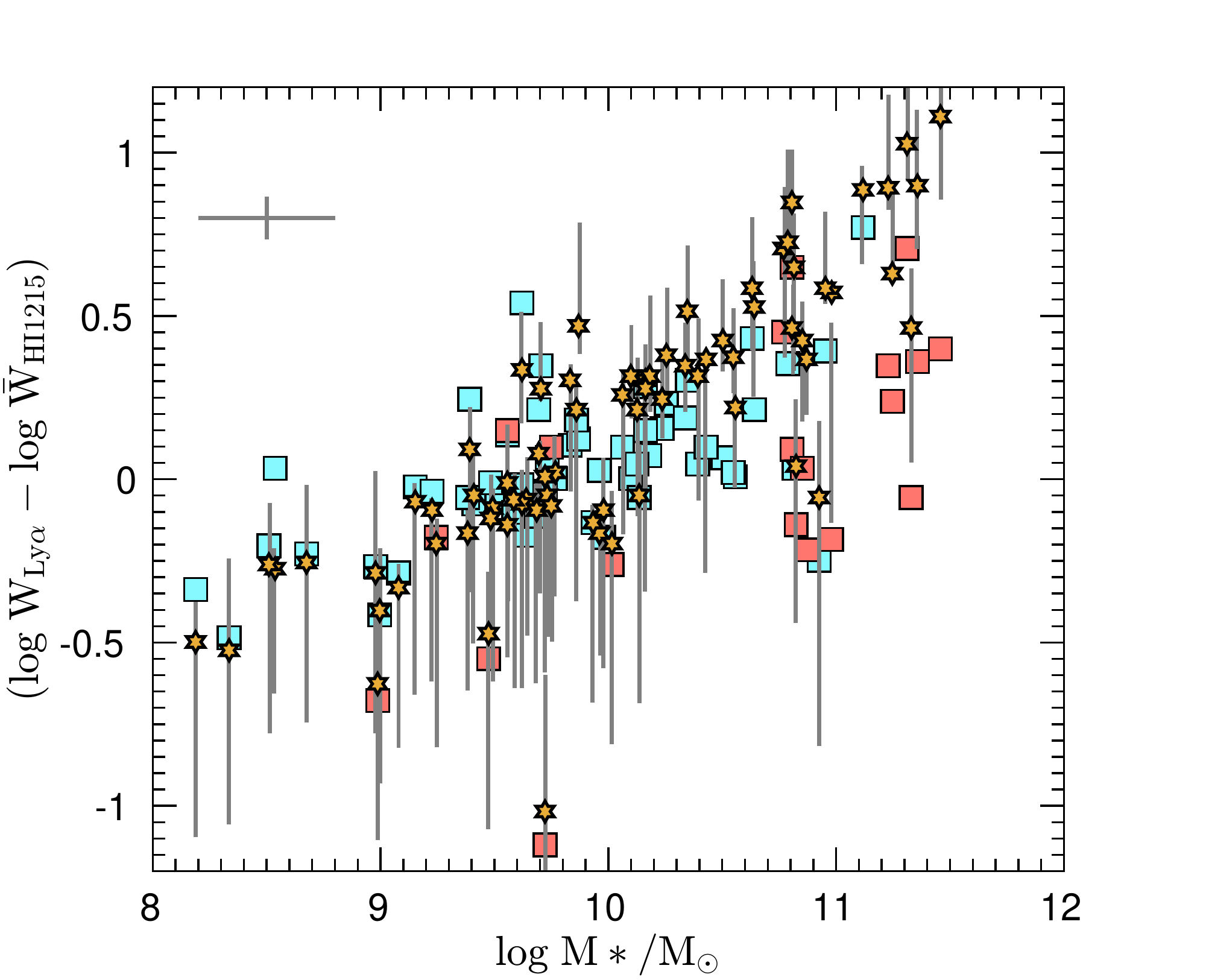}
\includegraphics[scale=0.435]{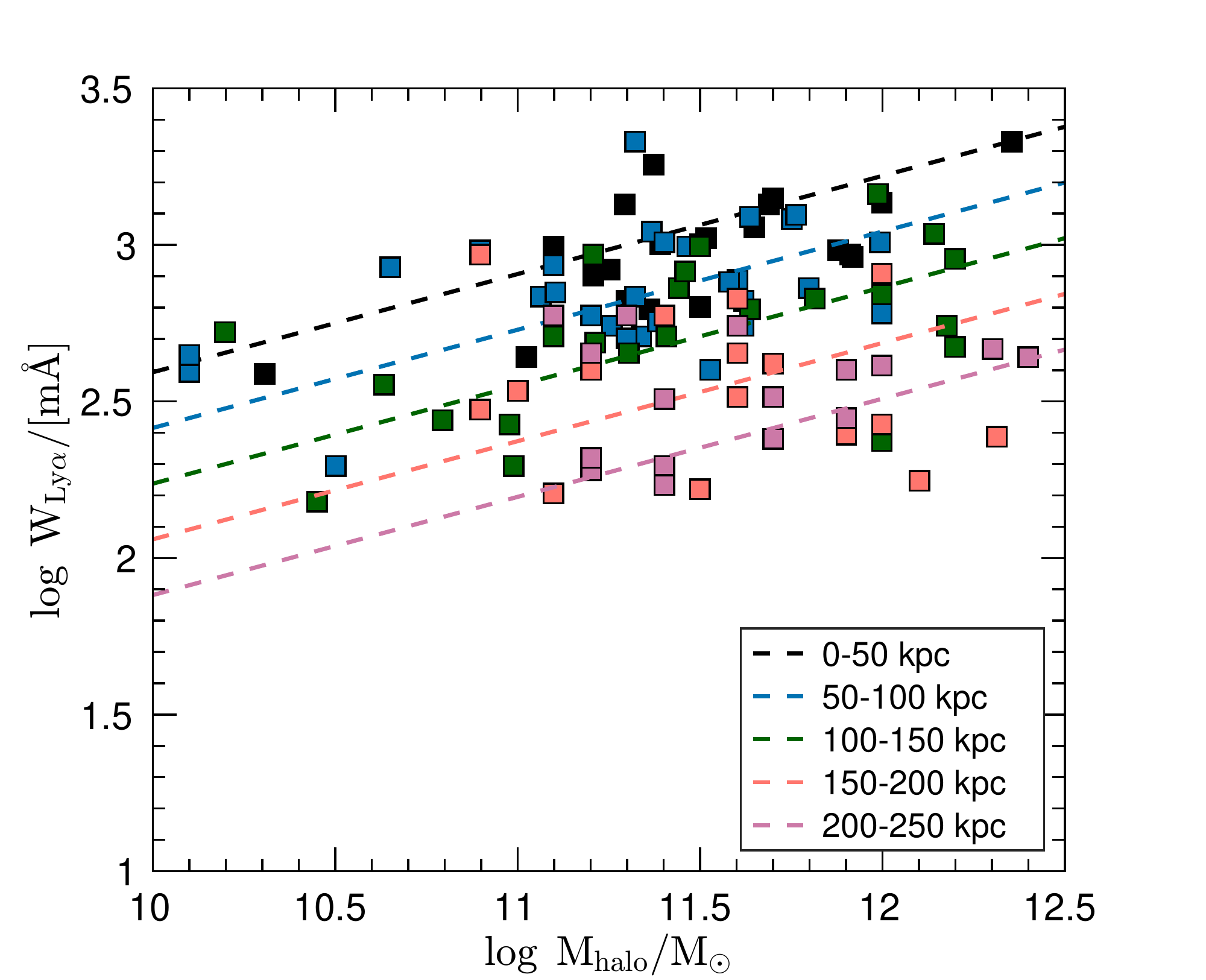}
\label{fig:toy_model}
\end{center} 
\caption{\textit{Left Panel:} Comparison of residual {\Lya} equivalent widths with stellar mass as in Figure \ref{fig:radial_profile}. The stars show the {\Lya} equivalent width residuals expected from a simple model, where {\Lya} equivalent width traces the same kinematics as their host galaxy's dark matter halos. \textit{Right Panel:} Halo mass estimates for {\Lya} absorption seen at different impact parameters are shown as dashed lines. The observed data points trace these halo mass estimates with reasonable accuracy. } 
\end{figure*}

We examine the hypothesis that \wlya\ traces the halo gravitational potential in a simple model. We make an ansatz that the trend of increase of {\Lya} residual equivalent width with increasing stellar mass is owing to higher velocity dispersion of the gas in higher mass dark matter halos. This ansatz is only valid for saturated absorbers in the flat portion of the curve of growth. Any unsaturated absorbers trace the underlying {\HI} column density rather than bulk gas kinematics and will add to the scatter seen in Figure \ref{fig:radial_profile} panels [b], [d], but we note that {\Lya} saturates at an equivalent
width of $\approx 0.1$\AA. Additionally, if a saturated {\Lya} line transitions to the damped part of the curve of
growth (e.g. for Damped {\Lya} absorbers), then the \wlya\ reflects the associated column density rather than the velocity dispersion. Such absorbers will also add to the scatter seen in Figure \ref{fig:radial_profile} panels [b], [d]. However, most of the {\HI} column densities associated with COS-Halos galaxies \citep{Prochaska2017}  are not high enough to produce strong damping wings, and no {\Lya} absorption profile associated with COS-Dwarfs galaxies exhibit damping wings (Bordoloi et. al. in prep). We assume that each halo possesses an NFW density profile \citep{Navarro1996} and a halo mass given by fitting abundance matching relations as described in \cite{Tumlinson2013}, and \cite{Bordoloi2014c}. For each galaxy, we compute the associated velocity dispersion ($\sigma_r$) at the given impact parameter ($R$) for its halo mass. For any {\Lya} absorber present in the CGM of that galaxy, the Doppler $b$ parameter is assumed to be $b \;=\; \sqrt{2} \sigma_r$. Then from a curve of growth, the expected {\Lya} equivalent width ($W_{model} (R,N,b)$) is computed, for a column density ($N$) and Doppler $b$ parameter. We stress that this model is not sensitive to column density for any saturated absorber in the flat portion of the curve of growth. We have well constrained {\HI} column density measurements for the COS-Halos galaxies  \citep{Prochaska2017}, and we use the apparent optical depth {\HI} column density estimates for COS-Dwarfs galaxies. The expected residual {\Lya} equivalent width values for this model is given as $\log \; W_{model} - \log \bar{W}_{HI 1215}$. Where $ \log \bar{W}_{HI 1215}$ is given by equation \ref{eqn:radial_profile}.  Figure \ref{fig:toy_model}, left panel shows the residual distribution for each sightline in this model (star symbols). The blue and red points are the {\Lya} residual distribution as shown in Figure \ref{fig:radial_profile}, panel [b]. The predictions from this simple model qualitatively reproduce the trend of enhanced \wlya\ with increasing stellar mass of the host galaxies.

To estimate the associated model uncertainties, we perform a suite of Monte Carlo simulations as follows. The main source of uncertainty involved in this simple model comes from the uncertainty in halo mass estimates,  uncertainty in {\HI} column density estimates, and uncertainty in $ \log \bar{W}_{HI 1215}$. These uncertainties propagate in a non-linear fashion. To account for all these uncertainties, we conservatively assume that the halo mass estimates have a 1$\sigma$ uncertainty of 1 dex associated with them. These are conservative numbers as typical abundance matching halo mass uncertainties are $\sim$ 0.2-0.3 dex \citep{Behroozi2010}. We also conservatively assume that the {\HI} column densities are uncertain by 2 dex. The uncertainty in $ \log \bar{W}_{HI 1215}$ is given in equation \ref{eqn:radial_profile}. The model was regenerated 10000 times, while randomly including these uncertainties. This gives a distribution of \wlya\ residuals for each galaxy. The width of this distribution is representative of model uncertainty associated with each galaxy. We take the width containing 68\% of the distribution as the estimation of model uncertainty. This is shown as gray error bars in Figure \ref{fig:toy_model}, left panel. It is clear that, even after accounting for conservative model uncertainties, the trend of increasing residual \wlya\  with stellar mass is well reproduced by this simple model. We conclude that the increase in \wlya\ with galaxy stellar mass at the inner CGM is primarily driven by the gas kinematics in the CGM. The cool gas clouds are plausibly bound to the dark matter halos of the host galaxies and are essentially tracing their dynamics.
 
Thus, {\Lya} absorption strength, impact parameter and the host dark matter halo mass form a fundamental plane. By fitting the observations of COS-Halos and COS-Dwarfs galaxies, we characterize this ``CGM fundamental plane'' as follows:
\begin{equation}
 \label{eqn:halo_profile}
 \log W^{\rm{f}}_{HI 1215} \; = \;  \alpha_{1} + \alpha_{2} R +  \alpha_{3} \log (M_{halo}/M_{\odot}).
 \end{equation}
The best fit coefficients with 95\% confidence bounds are found to be $\alpha_{1} =\;     -0.45 \;  (-1.7, 0.8)$, $\alpha_{2} \;= -0.004\;  (-0.005, -0.002)$, and $\alpha_{3} \;=\;  0.31 \; (0.2, 0.4)$, respectively. 

In this formulation, the dependence on $R$ is quite weak and halo mass is the dominant factor. As a consequence of this halo mass effect, this fit is reversible as an estimator for halo mass. The halo mass estimations derived from {\Lya} absorption equivalent widths at different impact parameters are shown as dashed lines in the right panel of Figure \ref{fig:toy_model}. The data points are a combined sample of archival surveys (as in Figure \ref{fig:radial_profile}, panel [c]), for which halo mass estimates are available. The points are colored coded to show different impact parameter ranges. The high and low impact parameter {\Lya} absorbers clearly segregate away from each other in equivalent width halo mass plane. This simple empirical fit given in Equation \ref{eqn:halo_profile} allows us to estimate halo masses from any CGM observations of isolated galaxies, simply knowing the equivalent width and impact parameter of a {\Lya} absorber.

The uncertainty on this halo mass estimator will come from the uncertainty introduced while fitting equation \ref{eqn:halo_profile}. Some of the uncertainty will come from the abundance matching halo masses used to create this fit. Another source of uncertainty is the intrinsic scatter seen around the surface fit (Figure \ref{fig:eqw_surf}, right panel). This may come from secondary macroscopic effects such as the environment in which the host galaxy resides \citep{Burchett2016}. Alternatively, higher column density {\HI} is also known to be associated with galaxy processes (e.g. accretion or extended disks for Damped {\Lya} absorbers), which will add to the uncertainty associated with this halo mass estimation method.

We can also asses the typical uncertainty associated with this halo mass estimator by using \wlya\ observations around two individual galaxies where multiple lines of sight pass through their CGM \citep{Keeney2013, Bowen2016}. \cite{Keeney2013} probe the CGM of a galaxy with halo mass $\log M_{halo} /M_{\odot} \sim$ 10.6 -- 11.4, at three impact parameters ($R$ = 74, 93 and 172 kpc). Using their \wlya\ values we derive a mean halo mass  $\log M_{halo} /M_{\odot} \approx {\rm 10.8 \pm 0.3}$.  Similarly,  \cite{Bowen2016} studied the CGM around a single galaxy (halo mass $\log M_{halo} /M_{\odot} \sim$ 12$\pm$0.2), using four sightlines ($R$ =48 -- 165 kpc). We again use their \wlya\ measurements to derive a mean halo mass $\log M_{halo} /M_{\odot} \approx {\rm 11.5 \pm 0.67}$.   Both these examples highlight the reasonable accuracy with which this new method can estimate halo mass of a system. Detailed future work will test and better constrain these uncertainties by comparing such observations and halo mass estimators with simulations. Such empirical scaling relations provide crucial independent constraints to estimate dark matter halo mass from observed gas dynamics.

In recent years it has been convincingly shown that there is a correlation between Mg~II equivalent width with halo and/or stellar mass \citep{Chen2010b,Churchill2013,Churchill2013b, Bordoloi2011, Zhu2014}. In Particular, \citet{Churchill2013,Churchill2013b} showed that, a strong correlation exists between Mg~II absorption equivalent width and virial radius normalized impact parameter of an absorber. 
Such a correlation  is also seen for other metal lines probing CGM of galaxies at different masses (e.g. \citealt{Prochaska2014,Liang2014, Bordoloi2014c, Johnson2015,Burchett2016}). \citet{Churchill2013b} found that, Mg~II absorption strength is determined by the virial mass of the host dark matter halo, while most of the Mg~II absorption resides at close impact parameter. However, at close impact parameters, strong Mg~II absorption also trace gas inflows and outflows  \citep{Bordoloi2011,Kacprzak2012,Bouche2013,Bordoloi2014d}. Such galactic origins of metal lines make it hard to use them as tracers of halo dynamics alone. Moreover, strong Mg~II is rarely seen at large impact parameters and so cannot be used to constrain halo dynamics further out. Incidence of other strong saturated metal absorption lines are generally low at $R > R_{vir}/2$ \citep{Bordoloi2014c,Liang2014,Johnson2015}.
Lastly, metallicity would certainly play a (non-linear) role in any such 
relation and may increase the observed scatter. 

We stress that this result is only valid for saturated {\Lya} absorbers on the flat part of curve of growth, observed within 200 kpc of their host galaxies. Moreover, this analysis, as noted in Section 2, is focused on relatively isolated galaxies.  Recent studies have shown that circumgalactic {\lya} absorption is very weak (or absent altogether) in clusters \citep{Burchett2017} and in some merging galaxies \citep{Johnson2014}. Equation \ref{eqn:halo_profile} may not apply to the CGM in those environments.

\section{Summary}
In this work, we present the variation of {\Lya} absorption strength in the circumgalactic medium of 85 galaxies from the COS-Halos and COS-Dwarfs surveys. All measurements are within 160 kpc from the host galaxies and the galaxies span a mass range of  $\rm{8 \leq log M*/ M_{\odot} \leq11.6}$. The main findings of this work are as follows.

\begin{itemize}

\item Across three decades in stellar mass, {\Lya} absorption with $W_{Ly\alpha} \geq$ 100 m{\AA} is ubiquitous around both star-forming (blue points) and passive (red points) galaxies out to 160 kpc.

\item After accounting for the radial dependence, the residuals of {\Lya} absorption show a strong trend of increasing {\Lya} equivalent width with increasing stellar mass of the host galaxy. A generalized Kendall's tau test rules out the null hypothesis that no correlation exists between {\Lya} absorption strength and the host galaxy stellar mass at  4.6$\sigma$ confidence (P = 5.02$\times 10^{-6}$).

\item This strong correlation persists, even after removing 9 high $R_{\rm vir}$ sightlines. For this sample, a generalized Kendall's tau test rules out the null hypothesis that no correlation exists between {\Lya} absorption strength and the host galaxy stellar mass at  $>$ 3.5$\sigma$ confidence (4.4$\times10^{-4}$).

\item The bi-variant dependence of {\Lya} equivalent widths with impact parameter and host galaxy stellar mass can be characterized by a surface fit given in equation \ref{eqn:surf_profile}. The surface can characterize the {\Lya} absorption strength around a star-forming galaxy with a scatter of $\sim$ 0.18 dex and that around a passive galaxy with a scatter of $\sim$ 0.29 dex. This simple empirical relation can be used to directly compare observations with theory and would help constrain numerical simulations.

\item The strong correlation of enhanced {\Lya} equivalent widths with increasing stellar mass of the host galaxy may be driven by dynamics of the gas. A simple model that assumes that the increase in equivalent width is proportional to the velocity dispersion of the host dark matter halo can explain the observed trend remarkably well. This suggests that such {\Lya} absorbers are typically bound to the dark matter halo of the host galaxy and are generally tracing the dynamics of the halo itself.

\item The observed {\Lya} absorption in the inner CGM traces the host dark matter halo properties. {\Lya} absorption strength, projected distance from the galaxy and mass of the host dark matter halo form a ``CGM fundamental plane'', which allows us to estimate the host dark matter halo mass from simple observations of {\Lya} absorption along a line of sight.

\end{itemize}

\acknowledgments
Support for this work was provided to R.B. by NASA through Hubble Fellowship grant \#51354. The COS-Halos and COS-Dwarfs datasets were collected under program IDs 11598 and 12248 respectively (PI Tumlinson). The fellowship and observing programs were all awarded by the Space Telescope Science Institute, which is operated by the Association of Universities for Research in Astronomy, Inc., for NASA, under contract NAS 5-26555. 

\bibliographystyle{yahapj}
\bibliography{references}

\end{document}